%% file: on-rk-theory-error-arXiv.tex
\documentclass[a4paper,11pt]{article}
\usepackage{pos}
\usepackage{xspace}
\usepackage{xcolor}

\title{On the $ R_{K} $ Theory Error}

\author*[a]{Saad Nabeebaccus}
\author[b]{Roman Zwicky}

\affiliation[a]{IJCLab, CNRS/IN2P3, Universit\'e Paris-Saclay,\\
 Orsay 91898, France}

\affiliation[b]{Higgs Centre for Theoretical Physics, School of Physics and Astronomy, University of Edinburgh,\\
	Edinburgh EH9 3JZ, Scotland}

\emailAdd{saad.nabeebaccus@ijclab.in2p3.fr}
\emailAdd{roman.zwicky@ed.ac.uk}

\input{my-commands.tex}

\abstract{To quantify the theory error on $R_K$, essentially means to quantify the 
uncertainty due to QED corrections since the latter breaks lepton flavour universality through  the lepton 
masses.  Since experiment uses photon shower programs, e.g. \texttt{PHOTOS}, 
 to capture QED effects, assessing the uncertainty  involves 
investigating effects not captured by the specific use of these tools. 
 This includes structure-dependent corrections, 
potentially large non-logarithmic terms and  charmonium resonances entering the lower bin by migration of radiation. We are able to close in on these loopholes. 
For example, using gauge invariance, we show that structure-dependent  
QED corrections do not lead to additional (sizeable) hard-collinear logs  of the form  ${\cal O}(\al) \ln m_\ell/m_B$.  
Hence, from the theory point of view $R_K$ is a safe observable.
}

\FullConference{%
  11th International Workshop on the CKM Unitarity Triangle (CKM2021)\\
  22-26 November 2021\\
  The University of Melbourne, Australia
}

\begin{document}
\maketitle

\section{Introduction}

In the Standard Model (SM),  lepton flavours couple to gauge bosons with the same coupling strength, giving rise to the concept of \textit{lepton flavour universality} (LFU). Thus, an important test of the SM is to consider \textit{LFU ratios}, and compare theory  with experiment.  
An example is $ R_{K} $ \cite{Hiller:2003js}, defined as the ratio of branching fractions of $ B\to K  \mu ^{+} \mu ^{-} $ and $ B\to K  e^{+} e^{-} $,
\begin{align}
	R_K\left[ q^2_{\mathrm{min}}, q^2_{\mathrm{max}} \right]=\frac{\int_{q^2_{\mathrm{min}}}^{q^2_{\mathrm{max}}}d q^2 \frac{d \Gamma \left( B\to K \mu^+ \mu^- \right)}{dq^2}}{\int_{q^2_{\mathrm{min}}}^{q^2_{\mathrm{max}}}d q^2 \frac{d \Gamma \left( B\to K e^+ e^- \right)}{dq^2}}\;,
\end{align}
in bins of $q^2= \left( \ell^{+}+\ell^{-} \right)^2  $ (the lepton pair momentum squared).  
The advantage of such ratios is that pure (non-perturbative) QCD corrections cancel as they are independent 
of the lepton flavour.

Since 2014, the  LHCb collaboration has been reporting discrepancies,
\begin{align}
\label{eq:RKex}
	R_K\left[ 1.1 \mathrm{GeV}^2,6 \mathrm{GeV}^2 \right]= 0.846^{+ 0.042 + 0.013}_{- 0.039 - 0.012}\;,
\end{align}
with the latest measurement \cite{LHCb:2021trn} (cf. \cite{BELLE:2019xld} for the Belle value compatible with theory albeit with significantly larger uncertainties),
whereas in theory
\begin{align}
\label{eq:RKth}
R_K\left[ 1.1 \mathrm{GeV}^2,6 \mathrm{GeV}^2 \right]\approx 1 + \Delta_{\textrm{QED}} \;,
\end{align}
with   $\Delta_{\textrm{QED}}$,  being specific QED corrections and  the subject of this proceeding. 
Ignoring $\Delta_{\textrm{QED}}$, \EQ\eqref{eq:RKex} implies a
  3.1 $  \sigma  $ deviation and, together with other deviations in $B$-physics, has 
  stimulated  some excitement in terms of model building in the 
  community.\footnote{
  Similar tensions between data and theory, albeit smaller statistical significance, 
have been measured  in the analogous quantities of $R_{K^{*0}}$~\cite{LHCb:2017avl}, $R_{K^{*+}}$ and $R_{K_S}$~\cite{LHCb:2021lvy}.  The Belle measurement of $R_{K^*}$ \cite{Belle:2019oag}
takes on the same role as for $R_K$.}

 Let us turn to QED. The crucial point is that the lepton masses do break LFU and that their
 scales are so different from the $b$ mass scale that they can give rise to significant corrections. 
 Concretely, the fine structure constant $ \frac{ \alpha }{ \pi }\approx 2 \cdot 10^{-3} $ is parametrically 
 enhanced by collinear logs by an order of magnitude  $\frac{\alpha}{\pi} \ln \frac{{m}_{e}}{m_B} \gtrsim 2$-$3\, \% $
 and with order one coefficient can amount up to $10\%$ \cite{BIP16,Isidori:2020acz} 
 or even $20\%$ towards the kinematic endpoint  \cite{Isidori:2022bzw}.
In practice the situation is more subtle as  experiment subtracts parts of 
$\Delta_{\textrm{QED}}$ 
from the result reported in \eqref{eq:RKex} by using the $\PHOTOS$ Monte Carlo tool. 
Thus, the title of this proceeding is slightly misleading and the
crucial question is how large the corrections are beyond the treatment in experiment.  
We assess the following three loopholes:
\begin{enumerate}
\item \emph{Hard-collinear logs beyond the point-like approximation} (structure-dependence). 
If that was the case then the uncertainty in the absence of a computation would make up $\ORD(10\%)$ and put the anomaly into question. 
\item  The \emph{charmonium resonances impact} on the $q^2$-bin \emph{below $6 \GeV^2$} by migration of radiation. 
In the case of the electrons, where a loose photon energy cut is used, 
a measured $q^2 \approx 6 \GeV^2$ probes much higher effective $q^2$ (cf. \TAB 1 in \cite{Isidori:2022bzw}). This is relevant since the charmonium modes are enhanced by $\ORD(10^4)$ with 
respect to the purely rare mode!
\item The $\PHOTOS$ program does \emph{not agree with the theory computation} in the point-like approximation.  As $\PHOTOS$ was tested in kaon physics and partly in $B$ physics this would constitute a surprise but in view of the stakes, it is better to test.  
\end{enumerate}
 We are able to give a largely positive answer, namely that $R_K$ is a rather safe observable 
 from the theory point of view.  In this proceeding we will largely focus on point 1, based on 
 \cite{Isidori:2020acz}, and try to summarise succinctly point 2 and 3, based on  \cite{Isidori:2022bzw}.  We note that reviewing the work in   \cite{Isidori:2020acz}, to which we first turn to, 
 will lay the groundwork for assessing  point 2. 

\section{QED effects in  $ \bar{B} \to \bar{K} \ell^+ \ell^- $ }

\subsection{Effective meson theory}

We start from an \textit{effective meson theory}, where the $ \bar{B} \to \bar{K} \ell_{1}\bar{\ell}_{2} $ decay is mediated by 
\begin{align}
{\cal L}^{\textrm {EFT}}_{\textrm{int}} = 
g_{ \mathrm{eff}}  \,   L^\mu V_\mu^{\textrm {EFT}} + \textrm{h.c.} \;,  
\label{eq:Lint_eff}
\end{align}
a lepton-hadron current interaction 
\begin{align}
L_\mu \equiv  \bar \ell_1 \gamma ^\mu( C_V + C_A  \gamma _5) \ell_2, \quad V_\mu^{\textrm {EFT}}  = \sum_{n \geq 0} \frac{f_\pm^{(n)}(0)}{n!}  
(-D^2)^n  [  (D_\mu B^\dagger)  K  \mp       B^\dagger( D_\mu  K)] \;,
\label{eq:ff}
\end{align}
where  the leptons $\lonetwo$ are kept generic, 
$ g_{ \mathrm{eff} } \equiv 2 \frac{G_F}{\sqrt{2}}   \lambda _{\textrm{CKM}} $,
$ C_{V(A)} \equiv -\frac{ \alpha }{2 \pi }C_{9(10)}$ are Wilson coefficients, $ f_\pm^{(n)}(0) $ represent the $ n^{ \mathrm{th} } $ derivative of the form factor $ f_\pm(q^{2}) $, evaluated at $ q^2=0 $, and $ D_{ \mu } $ is the covariant derivative, used to enforce  gauge invariance in its minimal form. 
This framework goes \emph{beyond} scalar QED (treating the mesons as point particles) in 
the expansion of the form factor.  The matching condition for the expansion is the reproduction of 
the leading oder (LO)  matrix element
\begin{eqnarray} 
\matel{\bar{K} }{ V_\mu }{\bar{B} }  = f_+(q^2) (p_B\pl p_K)_\mu + f_-(q^2) (p_B\mi p_K)_\mu   = \matel{\bar{K} }{ V^{\textrm {EFT}}_\mu }{\bar{B} }   +    {\cal O}(e) \;,\quad
\label{eq:match}
\end{eqnarray}
where $ V_{ \mu }\equiv  \bar{s}    \gamma _\mu  (1-\gamma_5) b  $.
Other than that,  mesons are treated as point-like particles. That is to say the   photon  itself  does not resolve the mesons.   
An important aspect to clarify are the kinematics as they control the IR-safety.

\subsection{Kinematics}

\begin{figure}[t]
		\vspace{-0.4cm}
	\centering
	\includegraphics[width=0.57\linewidth]{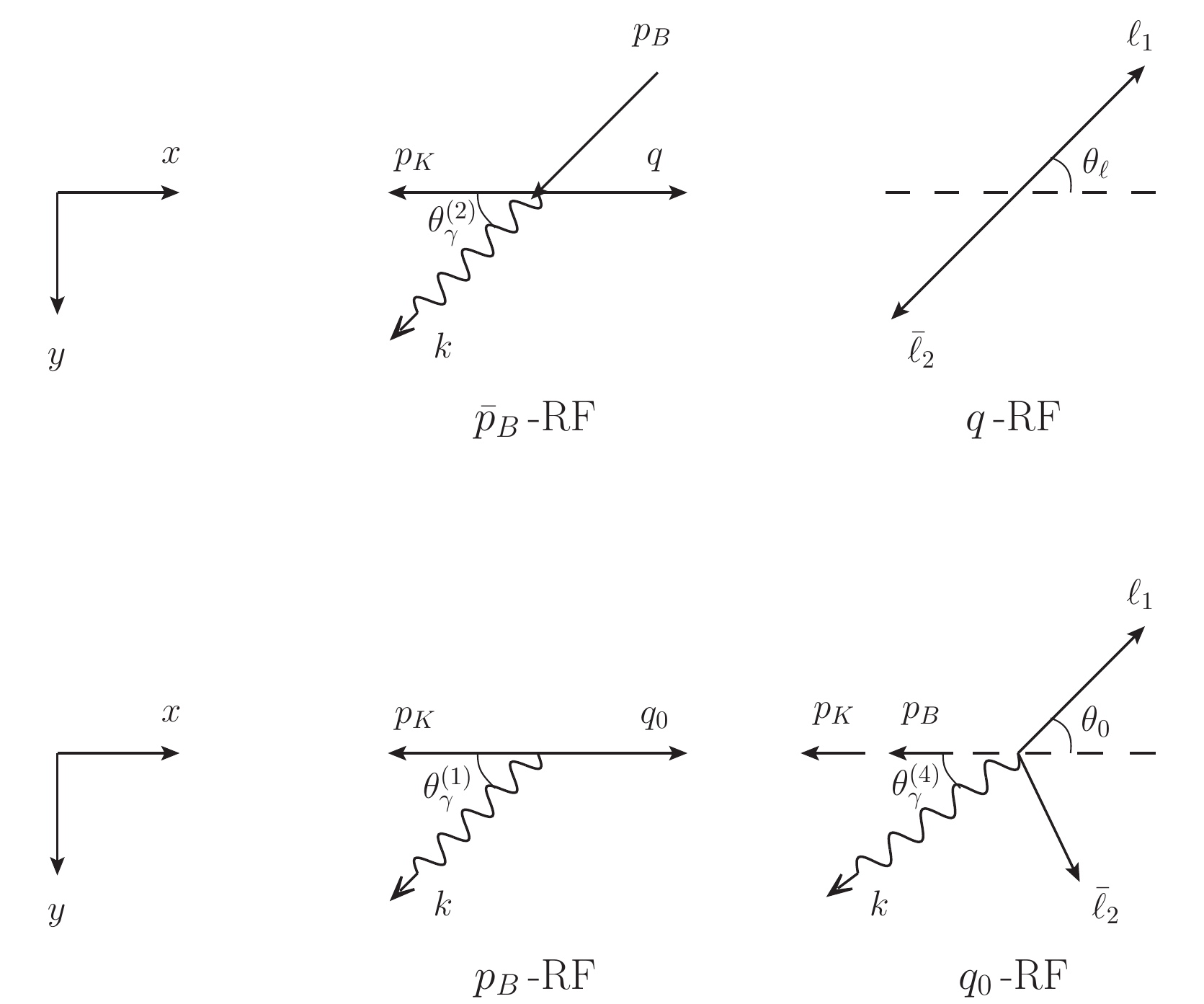}
	\vskip 0.5 true cm
	\caption{\label{fig:angles}
		\small Decay kinematics for the different RFs of interest.	 
		The dashed line corresponds to the decay axis, defined by the direction of the outgoing kaon.}
\end{figure}

We consider two sets of variables for the differential distribution of 
$\Bin(p_B) \to  \Kout(\pK)     \ell^+(\ell^+) \ell^-(\ell^-)   \gamma (k)$ process, 
assuming that radiation is not detected:  
\begin{equation} 
\label{eq:tr}
\{ \qSaa,  \claa \}   = \left\{  \begin{array}{lll}
q_\ell^2 = (\lone+\ltwo)^2,~ 
&  \cl =  - \left(\frac{\vec{\lone}\cdot \vec{p}_K}{ |\vec{\lone}| | \vec{p}_K| } \right)_{q-\textrm{RF}} ~
&  [\text{``Hadron collider'' variables}]~, \\
q^2_0 = (\pB-\pK)^2~, 
&  \clz=  - \left(\frac{\vec{\lone}\cdot \vec{p}_K}{ |\vec{\lone}| | \vec{p}_K| } \right)_{q_0 -\textrm{RF}}  ~
& \textrm{[``B-factory'' variables]} ~,    \end{array} \right. 
\end{equation}
where $q-\textrm{RF}$ and $q_0-\textrm{RF}$ denotes the rest frames  (RF) of 
\begin{equation}
q \equiv  \ell^+ + \ell^-  \;, \quad \qz \equiv  \pB-\pK = q + k \;,
\end{equation}
and $ \cl\equiv \cos  \Tl $, $ \clz\equiv \cos  \Tlz  $. 
This is illustrated in \FIG\ref{fig:angles}. 

For the real radiation, one needs to integrate over the photon momentum, and for this purpose, we define a cut-off for the photon energy (related to detector resolution) in a Lorentz invariant way\begin{align}
\pBbar^2 \equiv (p_B-k)^2 \, \geq \, m_B^2\left( 1- \des \right) \;,
\end{align}
using the experimentally reconstructed mass, $m_{B_{rec}}^2 = \pBbar^2 $, of the $ B $-meson.

\subsection{Computations and IR-safe Differential Variables  $ \{\qz^2,\clz\} $}

In computing the real and virtual parts, we employ phase space slicing to separate the IR sensitive terms 
into integrals that can be computed analytically. This leads to numerically stable cancellation of the
IR divergences (i.e. soft divergences regulated by the photon mass for example).

The differential rate is parameterised as follows
\begin{alignat}{2}
\label{eq:Delta}
&  d^2 \Gamma_{ \Bin\to \Kout \lone \bltwo}(\des )   &\;=\;& 
{d^2 \Gamma^{\LO}} +     \frac{ \alpha }{\pi} \sum_{i, j }   \hat{Q}_i \hat{Q}_j   
\left( {\cal H}_{ij} + {\cal F}^{(\AAA)}_{ij}(\des )   \right)    \, d  \qSaa d \claa   + {\cal O}( \alpha ^2)  \nonumber 
\\[0.1cm]
&  &\;=\;&  {d^2 \Gamma^{\LO}} \left[ 1 +  \Delta^{(\AAA)} (\qSaa,  \claa; \des)   \right]    \, d  \qSaa d \claa   + {\cal O}( \alpha ^2) \;,
\end{alignat}
where $ d^2 \Gamma^{\LO} $ corresponds to the LO differential rate,  
the indices $ i,j $  run  over all charged particles in the decay, 
and ${\cal H}$ and ${\cal F}$ stand for the virtual and real contributions respectively. 
The charges $ \hat{Q}_{i} $ are defined such that $ \sum_{i}\hat{Q}_{i}=0 $, cf. \cite{Isidori:2020acz}.

In order to separate IR sensitive regions of the real integration,  the \textit{two cutoff phase space slicing} procedure \cite{Harris:2001sx} is employed. It requires the introduction of two small unphysical parameters $ \{\Des,\Dec\} $ that allows the separation of the real part as follows
\begin{alignat}{2}
\label{eq:tactic}
& {\cal F}^{(a)}_{ij}(\des ) &\;=\;\;& \frac{d^2 \Gamma^{\LO}}{d q^2 d \cl}    \tF^\soft_{ij}  ( \Des ) +
\tF^\hca_{ij}(\underline{  \delta } ) +  \Delta {\cal F}^{(a)}_{ij}(  \underline{ \delta } )  \;,
\end{alignat}
where $   \underline{\delta}  =\{\Des,\Dec,\des\} $, and $ s $ and $ hc $ stand for `soft' and `hard-collinear'. By \textit{hard-collinear divergences}, we mean any photon emission with energy \textit{above} the soft cut-off $ \Des $, collinear to either lepton in the final state.

This procedure allows us to compute the IR sensitive real integrals $ \tF^\soft_{ij}$ and $ \tF^\hca_{ij} $ \textit{analytically} up to negligible terms of $\ORD(\omega_{c,s})$. 
The relevant dependence on the unphysical cut-offs is logarithmic and cancel in the sum of the three terms.  Powerlike corrections in $\omega_{s,c}$ are negligibly small.  
The \textit{numerical} integration in $ \Delta {\cal F}^{(a)}_{ij} $ is performed in the region where
\begin{align}
&\pBbar^2 \leq m_B^2 \left( 1-\Des \right), \quad k\!\cdot \! \lonetwo \geq \Dec m_B^2 \;.
\end{align}

We find that the soft and soft-collinear divergences cancel at the double differential level, independent of the choice of differential variables and  photon energy cut-off $ \des $. This is expected, by virtue of the KLN-theorem. The hard-collinear logs ($ \ln {\ml} $) are more interesting as 
the KLN-theorem guarantees their cancellation in the fully inclusive case (i.e. photon inclusive and integration over differential variables \eqref{eq:tr}).
Thus, the question is: does the cancellation survive  in any of the two sets of differential variables used?
 It turns out that $ \{\qz^2,\clz\} $ are the \textit{collinear-safe} variables, whereas for $ \{q^2,\cl\}, $ the hard-collinear logs do not cancel. This effect is at the heart 
of the $10$-$20\%$ QED corrections quoted in the introduction. Intuitively, the $ \{\qz^2,\clz\} $  variables 
are collinear-safe since for those, the $B$- and $K$-mesons can be thought of as one particle (with 4-momentum $ p_{B}-p_{K} $) and then 
its decay is analogous to a $Z \to \ell^+ \ell^-$ decay which is non-differential and IR finite 
according to  the KLN-theorem. 

\subsection{No new $\ORD(\al)$ hard-collinear logs at  structure-dependent level. }

A central result of our work is that we 
 were able to show, using \textit{gauge invariance} and the lepton equation of motion, that structure-dependent corrections (absent in our calculation, since we treated photon interactions with the mesons as \textit{scalar} interactions) do \textbf{not} contribute to hard-collinear logs $ \ln \ml $.
 Let us briefly sketch the argument , neglecting the lepton spin for the time being, 
 and refer the reader to 
 section 3.4 of \cite{Isidori:2020acz} for some detail.  
  The radiative amplitude can be decomposed into   
\begin{equation}
{\cal A} = \eps^* \Cdot (  {\cal A}_\ell + ({\cal A} -  {\cal A}_\ell)) \;, \qquad  \eps^* \Cdot {\cal A}_\ell  \propto \hat{Q}_{\ell} \frac{\eps^* \cdot \ell_\ell}{k \cdot \ell_\ell} \;,
\end{equation}
an eikonal part $ {\cal A}_\ell$ and a non-eikonal part.  Assuming the Feynman gauge, 
squaring the matrix element, summing over   polarisations  and integrating over the photon phase space one obtains the following three terms 
\begin{equation}
\int d \Phi_\ga \,  {\cal A}\Cdot {\cal A}^* = \int d \Phi_\ga ( ({\cal A} -  {\cal A}_\ell)\Cdot ({\cal A} -  {\cal A}_\ell)^* + 
2 \textrm{Re}[{\cal A}_\ell \Cdot   {\cal A}^*] -   {\cal A}_\ell \Cdot {\cal A}^*_\ell) \;.
\end{equation}
The first term is finite (in the $m_\ell \to 0$ limit)   in the collinear region (of the lepton). 
Crucially, in the second   we may use gauge invariance to establish proportionality to 
\begin{equation}
\ell_\ell \cdot {\cal A}  = k \cdot {\cal A}   +  \ORD(m_{\ell}^2) = \ORD(m_{\ell}^2) \;.
\end{equation}
From this we conclude that the second and third term give an expression of the form
\begin{equation}
2 \textrm{Re}[{\cal A}_\ell \Cdot   {\cal A}^*]  - {\cal A}_\ell \Cdot {\cal A}^*_\ell  
=  {\cal A}_\ell \Cdot {\cal A}^*_\ell + \dots \;,
\end{equation}
where the dots are free from hard-colinear logs. 
Finally, the $ {\cal A}_\ell \Cdot {\cal A}^*_\ell$-term  itself gives raise to the (universal)  hard-collinear  logs. 
Hence we learn that there cannot be any further hard-collinear logs in the structure-dependent part.  This is the case since structure-dependent contributions will just shift ${\cal A} \to {\cal A} + \de {\cal A}$ 
and  $k \cdot \de {\cal A} = 0$ by  gauge invariance.
When spin is considered the same conclusion holds \cite{Isidori:2020acz}.
This then clarifies point 1 raised in the introduction; there are no hard-collinear logs beyond the point-like approximation at $\ORD(\al)$.

\subsection{Results in terms of Plots}

\begin{figure}[h!]
	\includegraphics[width=0.42\linewidth]{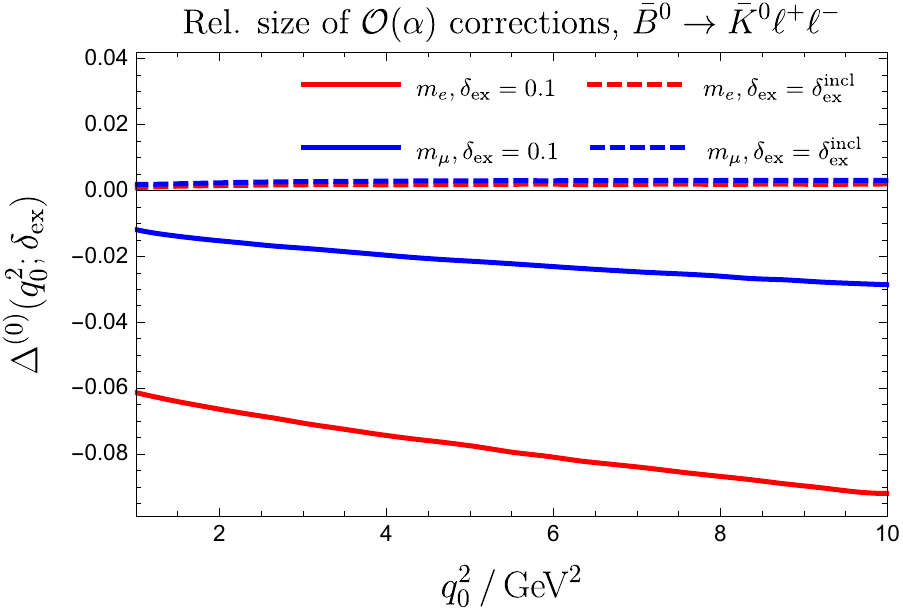}  
	\hfill
	\includegraphics[width=0.42\linewidth]{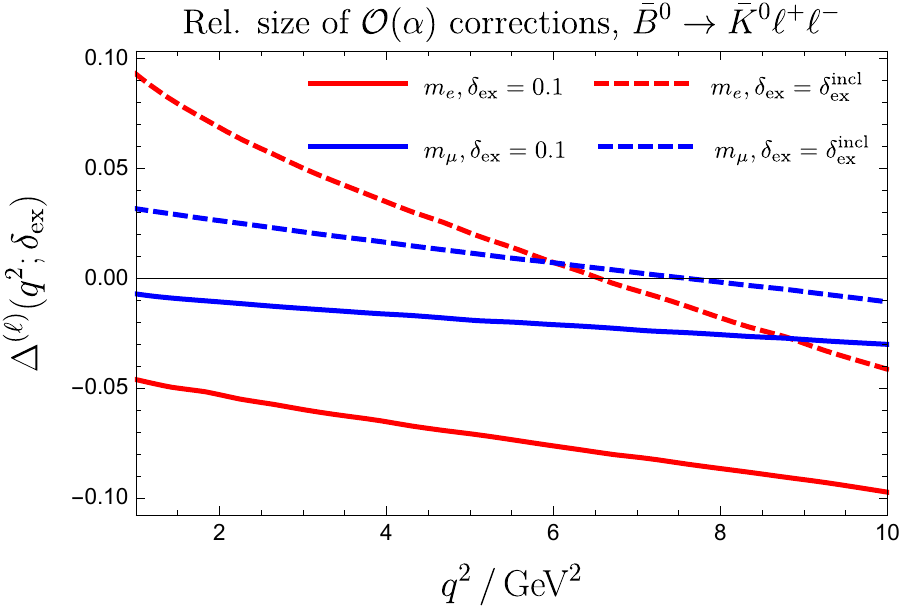}  
	\includegraphics[width=0.42\linewidth]{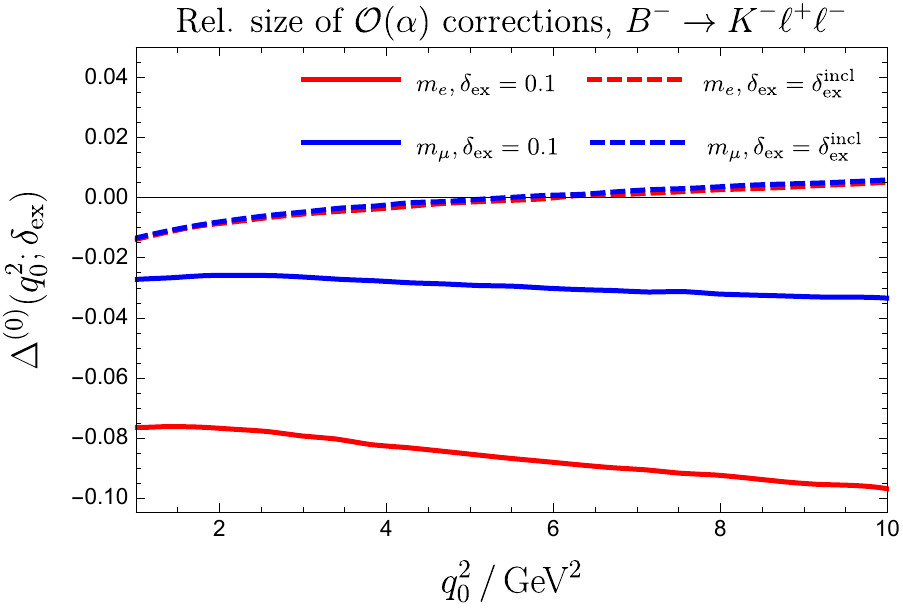}  
	\hfill
	\includegraphics[width=0.42\linewidth]{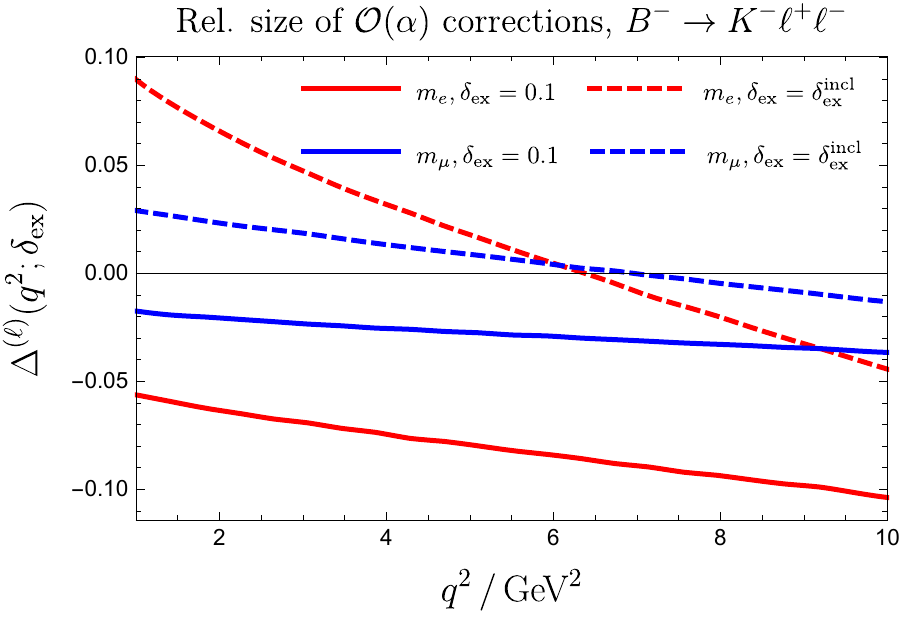}  
	\caption{\small The relative sizes of QED corrections are shown as a function of $ \qSaa $. The top and bottom plots represent the neutral and charged meson cases respectively. The left and right plots correspond to results differential in $ \qz^2 $ and $ q^2 $ respectively.}
		\label{fig:q2plots}
\end{figure}

We present the results as plots of QED corrections normalised with the LO differential rate, ie.
\begin{equation}
\Delta^{(\AAA)}(\qSaa;\des)  =   \left( \frac{d \Gamma^{\LO}}{ d\qSaa}\right)^{-1}  \frac{d\Gamma(\des)}{d \qSaa } \Big{|}_{ \alpha }   \;,  
\end{equation}
with the numerator and denominator separately integrated over the angular  variable $ \claa $ 
(defined in  \eqref{eq:tr}).
Our main plots are given in  \FIGs\ref{fig:q2plots} and \ref{fig:radmigration}, cf. \cite{Isidori:2020acz}
for further plots.

\FIG\ref{fig:q2plots}  is divided into the $q_0^2$- and $q^2$-variable on the left and right respectively, while bottom and top  correspond to neutral and charged mesons respectively.
For $q_0^2$  in the photon inclusive case (dashed lines and $\des= \desinc$),   
 the $\ln m_\ell$ terms cancel, as previously stated, and thus radiative corrections  are small $(\ORD(\frac{\al}{\pi}))$. 
 In the charged case, the ``hard-collinear'' logs of the kaon mass, $\ln m_K$, (bottom left) do seem to give rise to 
 a sizeable physical effect (up to $ \sim 2\% $).
  On the other hand, in the variable $ q^2 $, the hard-collinear logs do not cancel 
  at the differential level, and this explains why none of the lines remain close to zero. Of course, when one integrates over $ q^2 $ in the fully inclusive case, the hard-collinear logs have to cancel, which we have checked analytically.  This can be seen from the dashed lines on plots on the RHS in \FIG\ref{fig:q2plots} going from positive to negative values with increasing $ q^2 $. For a photon energy cut-off corresponding to
  $\des =0.1$, which is in between the values used for electrons and muons by LHCb, the QED effects
  are sizeable and are of course more pronounced for electrons as can be seen from the plots. 

Estimating  the QED correction to $ R_{K} $ based on our short distance analysis, 
using  photon cuts to emulate the  LHCb procedure \cite{LHCb:2021trn}, we find
\begin{equation}
\Delta_{\rm QED} R_K \approx   \left. \frac{ \Delta \Gamma_{K\mu\mu}}{ \Gamma_{K\mu\mu} } 
\right|^{m_B^{\textrm{rec}}= 5.175{\small \GeV}}_{q_0^2 \in [1,6]{\small \GeV}^2}
-  \left. \frac{ \Delta \Gamma_{K ee }}{ \Gamma_{Kee} } 
\right|^{m_B^{\textrm{rec}}= 4.88{\small \GeV}}_{q_0^2 \in [1,6]{\small \GeV}^2}
\approx +1.7\%\;,
\end{equation}
which is accidentally small due to the cuts in use. 
In \cite{BIP16}, a correction of $ \Delta_{\rm QED} R_K \approx  3\% $ was reported, where a tight angle cut was applied, in addition to photon energy cuts. 

\subsection{Migration of the charmonium resonances}

\begin{figure}[t]
	\centering
	\includegraphics[width=0.44\linewidth]{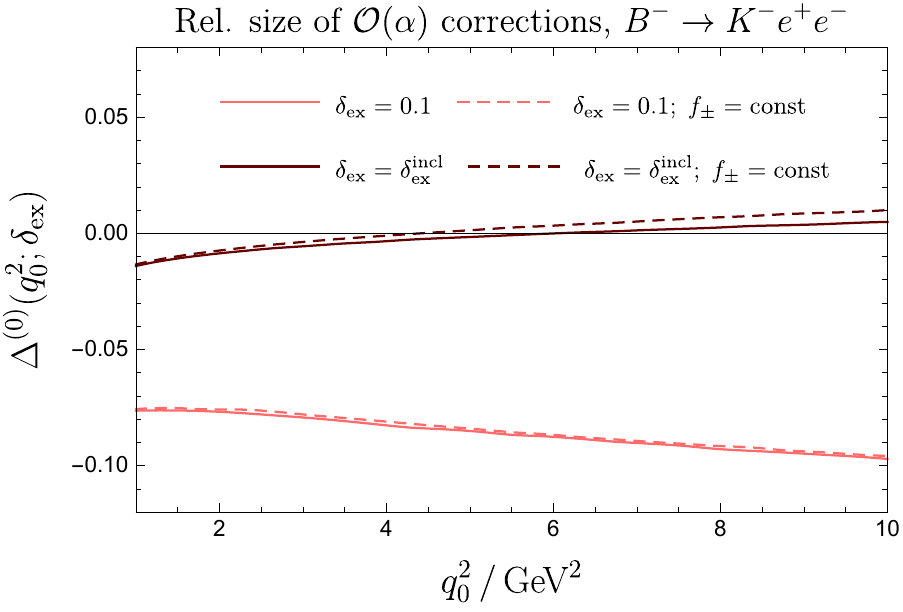}  
	\hfill 
	\includegraphics[width=0.44\linewidth]{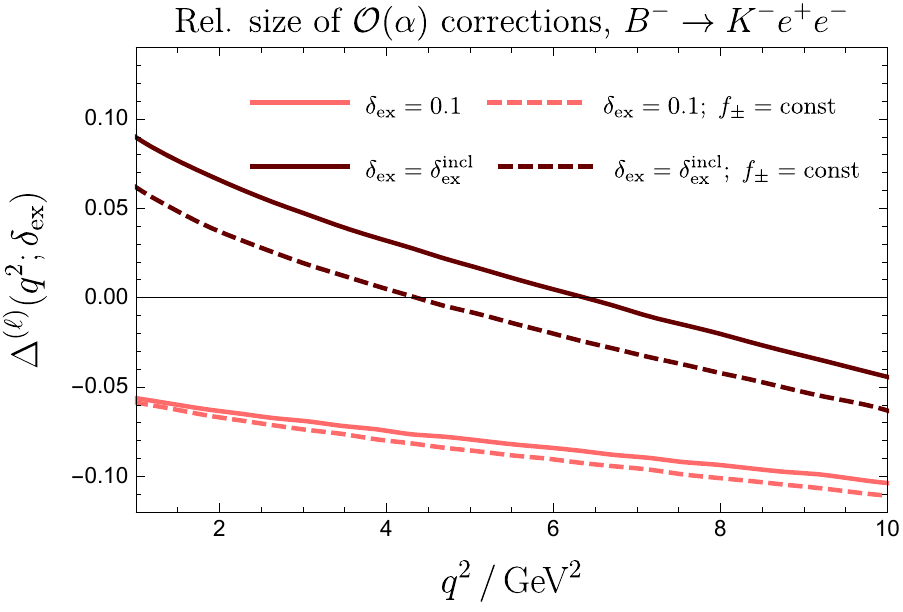}  
	\caption{\small The effects of migration of radiation are shown. The left and right plots represent relative QED corrections in $ \qz^2 $ and $ q^2 $ respectively. Dashed lines correspond to a constant form factor, while the solid line includes the first derivative term of the form factor 
	expansion (see  \EQ\eqref{eq:ff}).}
	\label{fig:radmigration}
\end{figure}

The \textit{migration of radiation} effect needs to be properly assessed, in view of the charmonium resonances. 
The  migration effect is illustrated in \FIG \ref{fig:radmigration} 
 by choosing different shapes of $q^2$-dependence for the form factors. 
Specifically,  we do 
so by plotting  the constant part of the form factors (dashed lines) versus 
the $q^2$-expanded form factor (solid line)  to linear order  (cf. \EQ\eqref{eq:ff}).
 The pale pink colour corresponds to a photon cut-off of $ \des=0.1 $, while the dark red colour corresponds to the fully inclusive case. It is found  that the effect of the form factors is small when differential in $ \qz^2 $. However, for $q^2$    the size of the relative corrections are \textit{significantly} affected by the form factors (cf. \FIG\ref{fig:radmigration}). The effects are larger when one is more photon inclusive. This is due to the fact that, for the $ q^2 $ distribution, a fixed value of $ q^2 $ probes \textit{higher} values of $ \qz^2 $, and the looser the photon energy cut-off, the wider is the range of $ \qz^2 $ that is probed (in fact, in the fully-inclusive case, the \textit{entire} spectrum is probed).

In view of this understanding let us discuss the charmonium resonances.
Currently, the LHCb  experiment is treating the resonant part  as a separate process to the rare one. 
This gives rise to the following 
two possible issues:
\begin{itemize}
\item [A.] Interference effects, which are neglected due the separate treatment,  could be large. 
\item [B.] The very specific treatment of the charmonium resonances is sizeable, due to the 
$\ORD(10^4)$-enhancement, 
in the case of the loose electron-cut.\footnote{It should be emphasised that these effects are of course less severe at Belle II since there $q_0^2$ can be measured, rather than only $q^2$ in the case of 
LHCb, as the initial $B$-meson momentum is known (to sufficient accuracy.)}
\end{itemize}
Fortunately, neither is  the case \cite{Isidori:2022bzw}.  With regard to point  A.  we refer the reader to 
\FIG 3,4 in \cite{Isidori:2022bzw}.\footnote{Note that due to Monte Carlo sampling efficiency  issues the electron mass is taken 
to be ten times its actual value but the conclusion are unchanged for the actual electron mass.}  To assess 
point B.  we used a splitting function approach, which reproduces our result numerically well. 
From \FIG 5,6  \cite{Isidori:2022bzw} one can infer that for $q^2 < 6 \GeV^2$ the impact is not too large (and we refer 
the reader to further comments at the end of \SEC 4.3 for the robustness of this statement). 
Note that for  $q^2 = 7 \GeV^2$ the effect is out of control. Hence  the electron 
cut is just tight enough to keep this effect under control. 
This then clarifies point 2; the loose electron cut is just tight enough to maintain the uncertainty, 
due to the charmonium resonances ,well below the statistical experimental error.

\subsection{Comparison with \PHOTOS}

In view of the importance of $R_K$  it seems imperative to scrutinies \PHOTOS versus 
our theory computation.  More accurately, a purpose built Monte Carlo program was developed  in \cite{Isidori:2022bzw}, based on our theory computation, using a hit-or-miss algorithm.  
The comparison between \PHOTOS and our Monte Carlo, using the same $\ORD(10^6)$ event samples,  
was made in \FIGs 1,2 in \cite{Isidori:2022bzw}, and excellent agreement was found!\footnote{\PHOTOS employs  a splitting function approach which captures the leading logs of the point-like particles and resums the soft logs to all orders \`a la YFS. The virtual corrections are indirectly inferred from the KLN-theorem which again captures the leading logs. Hence 
agreement with \PHOTOS was to be expected since the splitting function approach reproduces 
our results too very reasonable accuracy.} 
This then clarifies point 3; \PHOTOS captures QED effects due to the point-like approximation 
well below the percent level.

\section{Conclusion}

Our computational results, in the point-like approximation,  
show that it is  important to properly take into account QED corrections, as these are enhanced by hard-collinear logs $\ORD(\al) \ln \frac{m_\ell}{m_B}$.  Soft and soft-collinear divergences are universal, cancel at the differential level and  resurface as $\ORD(\al) \ln \des$  and $\ORD(\al) \ln \des \ln \frac{m_\ell}{m_B}$ effects.
  Hard-collinear divergences are more interesting as they are higher in energy and could probe the mesonic structure. Moreover, they may not cancel, depending on the choice of differential variables (even in the fully photon inclusive case).

However, we were able to show that all three loopholes raised in the introduction do not 
enhance the uncertainty.
In particular  gauge invariance allowed us to show that  hard-collinear logs are absent at the structure-dependent level.\footnote{Since the $K$-meson is not a point-like particle 
sizeable $\ln m_K$-effects can be expected to be present at the structure-dependent level 
which are however LFU (provided photon energy cut is the same for electrons and muons) 
but possibly relevant for the precision extraction of CKM elements. 
In $B$-physics the methods to capture structure-dependence 
necessitate the introduction of new gauge invariant interpolating operators  \cite{Nabeebaccus:2022jhu} 
and or new gauge invariant 
distribution amplitudes  \cite{Beneke:2021pkl}.}
The migration of charmonium resonances is \ emph{just} under control for $q^2 = 6 \GeV^2$. 
Actual comparison between a purpose built Monte Carlo tool and $\PHOTOS$ shows excellent agreement 
and this provides further evidence that  $\PHOTOS$ is a reliable tool in $B$-physics.  
Hence \emph{the theory uncertainty} of  $R_K$ is, according to our findings,  small and well under control. The same applies to other LFU ratios such as $R_{K^*}$ or the even less problematic $R_{D^{(*)}}$ 
as they are of the semileptonic type free from low lying resonances.

\bibliographystyle{JHEP}
\bibliography{masterrefs.bib}

\end{document}

%% file: my-commands.tex
\newcommand{\al}{\alpha}

\newcommand{\ga}{\gamma}
\newcommand{\de}{\delta}

\newcommand{\eps}{\epsilon}

\newcommand{\TAB}{Tab.~}
\newcommand{\FIG}{Fig.~}
\newcommand{\FIGs}{Figs.~}
\newcommand{\SEC}{Sec.~}

\newcommand{\EQ}{Eq.~}

\newcommand{\matel}[3]{\langle #1|#2|#3\rangle}

\newcommand{\GeV}{\,\mbox{GeV}}

\newcommand{\ORD}{{\cal O}}

\newcommand{\Cdot}{\!\cdot \!}
\newcommand{\mi}{\!-\!}
\newcommand{\pl}{\!+\!}

\newcommand{\Bin}{{\bar{B}}}
\newcommand{\Kout}{{\bar{K}}}

\newcommand{\soft}{{(s)}}

\newcommand{\hca}{{(hc)(a)}} 
 
\newcommand{\tF}{{\tilde{\cal F}} }

\newcommand{\des}{\de_{\textrm{ex}}} 
\newcommand{\desinc}{\de_{\textrm{ex}}^{\textrm{inc}}}

\newcommand{\AAA}{a }

\newcommand{\Des}{\omega_s}

\newcommand{\Dec}{\omega_c}

\newcommand{\LO}{\,  \mathrm{{ LO}}}

\newcommand{\qSaa}{q^2_{\AAA}}

\newcommand{\lonetwo}{\ell_{1,2}}

\newcommand{\lone}{\ell_1}
\newcommand{\ltwo}{\ell_2}

\newcommand{\bltwo}{\bar{\ell}_2}
\newcommand{\pK}{p_K}
\newcommand{\pB}{p_B}

\newcommand{\ml}{m_{\ell}}

\newcommand{\pBbar}{\bar{p}_B}

\newcommand{\Tl}{\theta_{\ell}}

\newcommand{\qz}{q_0}

\newcommand{\Tlz}{\theta_0}

\newcommand{\claa}{c_{\AAA}}

\newcommand{\cl}{c_{\ell}}

\newcommand{\clz}{c_{0}}

\newcommand{\PHOTOS}{\texttt{PHOTOS}\xspace}

\definecolor{lightblue}{rgb}{0.39, 0.58, 1.00} \definecolor{lightgreen}{rgb}{0.1, 0.73, 0.33}